\documentclass[aps,prl,twocolumn,showpacs,amsmath,amssymb,superscriptaddress,floatfix,nofootinbib]{revtex4-1}

\usepackage{color}         
\usepackage{graphics}      
\usepackage[pdftex]{graphicx}      
\usepackage{bm}
\usepackage{natbib}            
\usepackage[colorlinks,plainpages=false,linkcolor=blue,urlcolor=blue,citecolor=blue,pdfpagemode=UseNone,pdfstartview=FitBH]{hyperref}
\usepackage{float}

\newcommand{\lno}{LaNiO$_3$ }
\newcommand{\dso}{DyScO$_3$ }
\newcommand{\qhalf}{($\frac{1}{2}$ 0 $\frac{1}{2}$) }
\newcommand{\qhalfneg}{($\frac{1}{2}$ 0 $\overline{\frac{1}{2}}$) }

\begin{document}
\title{Transfer of magnetic order and anisotropy \\ through epitaxial integration of 3$d$ and 4$f$ spin systems}

\author{M.~Bluschke}
\affiliation{Max Planck Institute for Solid State Research, Heisenbergstr. 1, 70569 Stuttgart, Germany}
\affiliation{Helmholtz-Zentrum Berlin f\"{u}r Materialien und Energie, Wilhelm-Conrad-R\"{o}ntgen-Campus BESSY II, Albert-Einstein-Str. 15, 12489 Berlin, Germany}

\author{A.~Frano}
\affiliation{Department of Physics, University of California, Berkeley, California 94720, USA}
\affiliation{Advanced Light Source, Lawrence Berkeley National Laboratory, Berkeley, California 94720, USA}

\author{E.~Schierle}
\affiliation{Helmholtz-Zentrum Berlin f\"{u}r Materialien und Energie, Wilhelm-Conrad-R\"{o}ntgen-Campus BESSY II, Albert-Einstein-Str. 15, 12489 Berlin, Germany}

\author{M.~Minola}
\affiliation{Max Planck Institute for Solid State Research, Heisenbergstr. 1, 70569 Stuttgart, Germany}

\author{M.~Hepting}
\affiliation{Max Planck Institute for Solid State Research, Heisenbergstr. 1, 70569 Stuttgart, Germany}

\author{G.~Christiani}
\affiliation{Max Planck Institute for Solid State Research, Heisenbergstr. 1, 70569 Stuttgart, Germany}

\author{G.~Logvenov}
\affiliation{Max Planck Institute for Solid State Research, Heisenbergstr. 1, 70569 Stuttgart, Germany}

\author{E.~Weschke}
\affiliation{Helmholtz-Zentrum Berlin f\"{u}r Materialien und Energie, Wilhelm-Conrad-R\"{o}ntgen-Campus BESSY II, Albert-Einstein-Str. 15, 12489 Berlin, Germany}

\author{E.~Benckiser}
\affiliation{Max Planck Institute for Solid State Research, Heisenbergstr. 1, 70569 Stuttgart, Germany}

\author{B.~Keimer}
\email[]{B.Keimer@fkf.mpg.de}
\affiliation{Max Planck Institute for Solid State Research, Heisenbergstr. 1, 70569 Stuttgart, Germany}



\begin{abstract}
Resonant x-ray scattering at the Dy $M_5$ and Ni $L_3$ absorption edges was used to probe the temperature and magnetic field dependence of magnetic order in epitaxial LaNiO$_3$-DyScO$_3$ superlattices. For superlattices with 2 unit cell thick \lno layers, a commensurate spiral state develops in the Ni spin system below 100 K. Upon cooling below $T_{ind} = 18$ K, Dy-Ni exchange interactions across the LaNiO$_3$-DyScO$_3$ interfaces induce collinear magnetic order of interfacial Dy moments as well as a reorientation of the Ni spins to a direction dictated by the strong magneto-crystalline anisotropy of Dy. This transition is reversible by an external magnetic field of 3 T. Tailored exchange interactions between rare-earth and transition-metal ions thus open up new perspectives for the manipulation of spin structures in metal-oxide heterostructures and devices.
\end{abstract}

\maketitle

Exchange interactions between the localized $4f$ electrons in rare earths and the more itinerant $3d$ electrons in transition metals are of key importance in a wide spectrum of materials ranging from permanent magnets \cite{McCallum} to multiferroics \cite{Matsukura}.  The transition-metal subsystems in such materials typically exhibit magnetic order at high temperatures. Coupling to the rare-earth moments with large spin-orbit coupling pins the order parameter to the crystal lattice and greatly enhances the magneto-crystalline anisotropy that is central to the functionality of many magnetic materials. Progress in understanding and tailoring $3d-4f$ interactions is currently limited by the ability to efficiently integrate both subsystems in multinary compounds \cite{McCallum,Matsukura}.

Metal-oxide heterostructures and superlattices have recently emerged as a powerful platform for the investigation and control of spin, charge, and orbital order \cite{Keimer_oxideinterfaces}. In field-effect devices built from such structures, gate electric fields can efficiently counteract built-in polarization fields and thus elicit large changes in the carrier density and transport properties. Manipulation of antiferromagnetic structures through external magnetic fields would open up many additional perspectives for electronic devices, but the large fields required to counteract the exchange interactions between transition metal ions prohibit most experimental investigations along these lines. Here we show that $3d-4f$ interactions in metal-oxide heterostructures can pin the orientation of the 3$d$ moments to directions determined by the anisotropy of the rare earth ions across the epitaxial interface. As a function of temperature, this pinning results in a spin reorientation transition within the transition metal layer. This transition is fully reversible by magnetic fields of a few Tesla, which couple to the 3$d$ moments indirectly via the large Dy moments. Rare-earth moments can thus serve as ``anchors'' for the magnetic-field-induced manipulation of magnetic order in metal-oxide heterostructures. In this way, the $3d$-electron system is disturbed only at the boundary, contrary to bulk multinary compounds where $4f$ elements inextricably influence both the magnetic interactions and the lattice structure.

The multilayer system we have chosen comprises two compounds, LaNiO$_3$ and DyScO$_3$, that have already been studied individually as components of metal-oxide heterostructures. Both compounds share the same pseudocubic perovskite structure with similar lattice parameters, and are therefore well suited for epitaxial integration. LaNiO$_3$ is metallic and paramagnetic in its bulk form \cite{goodenough_raccah_1965_perovskite_oxides, Demazeau_1971_trivalent_nickelates, LACORRE_1991_JSSC}, but non-collinear antiferromagnetism of the Ni ions appears below $T_{N} \sim 100 $ K when 2 pseudocubic unit cells (u.c.) thin LaNiO$_3$ layers are confined between insulators \cite{Sasha,Frano_SDW_nickelates}. DyScO$_3$ is commonly used as a substrate for metal-oxide heterostructures. It is a paramagnetic insulator and is known to exhibit magnetic order driven by Dy-Dy exchange interactions below 3.1 K \cite{DSO_Dy_order,Raekers_Scandates}, but the magnetic structure and anisotropy have not yet been investigated in detail. The trivalent La and Sc ions in the compounds constituting the superlattice have closed electronic shells and are nonmagnetic.

We have used pulsed laser deposition to grow superlattices of \lno and DyScO$_3$, using a previously established protocol \cite{Supplementary_Material}. Three different superlattices have been studied: a) 2 u.c. \lno + 2 u.c. \dso ($2\vert\vert2$) repeated 15 times, b) 2 u.c. \lno + 4 u.c. \dso ($2\vert\vert4$) repeated 11 times, and c) 4 u.c. \lno + 4 u.c. \dso ($4\vert\vert4$) repeated 12 times. The superlattices were grown on \dso substrates oriented in the (110) direction of their orthorhombic structure (space group $Pbnm$). The $Pbnm$ setting of the substrate is used throughout this article to index the reciprocal space of the superlattice. To develop a detailed, element- and layer-specific description of the magnetic structure of the superlattices, we have used a combination of magnetometry and resonant x-ray scattering (RXS) \cite{Fink_REXS_review}. The RXS experiments were performed at the BESSY II undulator beamline UE46-PGM1 with photon energies tuned to the Ni $L_3$ edge (photon energy $E = 853.4$ eV) or to the Dy $M_5$ edge ($E = 1293.8$ eV), using a three-circle ultra-high-vacuum diffractometer. The magnetization was measured with a Quantum Design VSM magnetometer.

\begin{figure}[ht]
\includegraphics[width=8cm]{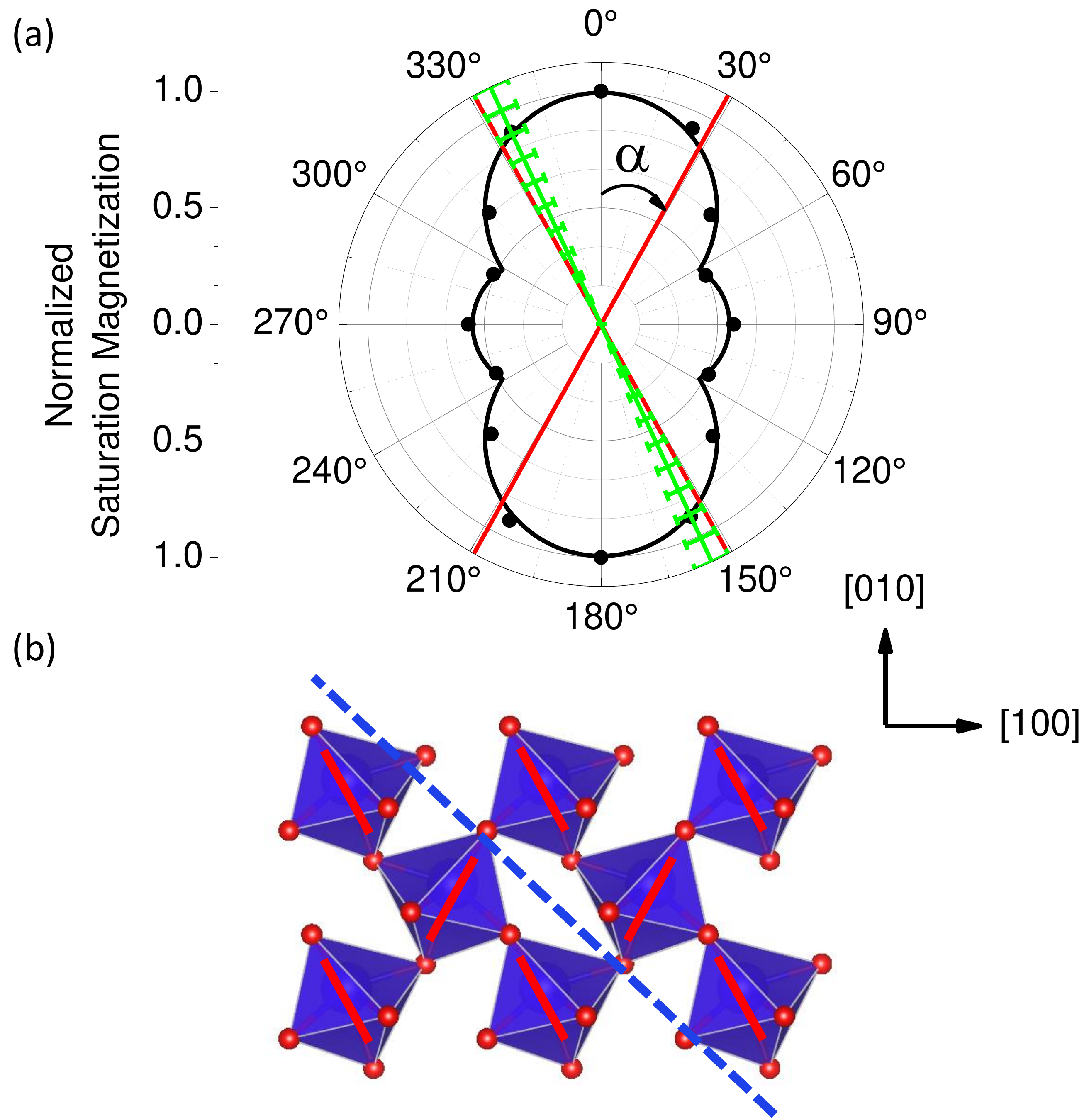}
\caption{(a) Angular dependence within the $a$-$b$ plane of the macroscopic saturation magnetization in bulk DyScO$_3$. The black line is the result of the Ising model which has been fitted to the data to determine the orientation of the two Ising axes (red lines). (b) Pictorial representation of the DyScO$_3$ structure \cite{Momma:db5098} with the two Ising axes (only Dy and its nearest coordinating oxygens are shown). One of the Ising axes agrees closely with the axis of the collinear Dy antiferromagnetism determined by RXS in the superlattices (solid green line in panel (a)). The dashed blue line is in the plane of the (110) oriented superlattice interface.
\label{fig:bulk_DSO_magnetometry}}
\end{figure}

Before discussing the results of the RXS experiments, we describe measurements of the macroscopic magnetization of bulk DyScO$_3$ which will be important to understand the x-ray data. Figure \ref{fig:bulk_DSO_magnetometry}(a) shows the saturation magnetization measured at $T =$ 2 K as a function of angle in the $a$-$b$ plane, along with the results of an Ising model inspired by prior work on the isostructural compounds DyAlO$_3$, TbAlO$_3$, and DyCoO$_3$ \cite{Holmes_DyAlO3_Metamagnetic_Behaviour, Holmes_TbAlO3, Velleaud_1975_DyScO3}. The magnetic Dy$^{3+}$ ions in the orthorhombically distorted perovskite structure are coordinated by 12 neighboring oxygens, which produce a crystal field of point-group symmetry $C_{1h}$. The crystal field splits the $^6H_{15/2}$ manifold of the free ion into 8 Kramers doublets, with a magnetic moment of 10 $\mu_B$ in the $M_J = \pm 15/2$ ground-state doublet. The low-symmetry crystal field produces a strong Ising anisotropy, with two inequivalent Ising axes (related by a mirror reflection in the $b$-$c$ plane) that alternate from site to site (Fig. \ref{fig:bulk_DSO_magnetometry}(b)) \cite{schuchert_1969_DyAlO3, Holmes_DyAlO3_Metamagnetic_Behaviour}. Since both Ising axes lie in the $a$-$b$ plane, the angular dependence of the saturation magnetization in this plane can be described by the sum of the absolute values of two cosine functions $\propto \vert cos(\alpha) \vert + \vert cos(-\alpha) \vert$ i.e., the projection of the full moments onto the direction of the applied field (black line in Fig. \ref{fig:bulk_DSO_magnetometry}(a)). The Ising axes are found to lie at $\alpha=29^{\circ}$ from the $b$ axis. From the anisotropy of the Dy moments' saturation (measured up to 7 T) it can be concluded that in this field and temperature regime the exchange energy $<<$ Zeeman energy $<<$ crystal field and spin-orbit energies.

\begin{figure*}[htb]
\includegraphics[width=17cm]{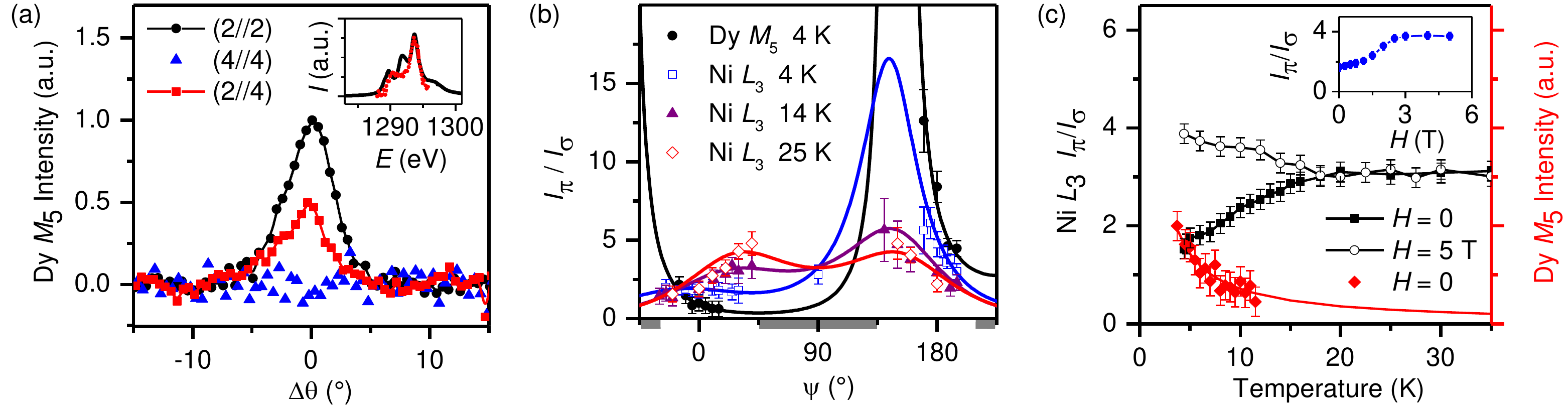}
\caption{(a) Rocking curves around the  $\mathbf{q}=$ \qhalf magnetic Bragg peak for various superlattices, with intensity normalized to the number of \dso layers in the superlattice. Measurements were performed at $T = 4$ K and with $\pi$-polarized photons tuned to the Dy $M_5$ edge. The inset shows the photon energy dependence of the $\mathbf{q}=$ \qhalf magnetic Bragg peak intensity across the Dy $M_5$ edge (red circles) compared with the energy dependent x-ray absorption measured in the total electron yield mode (black line). (b) Measured (points) and calculated (lines) azimuthal-angle ($\psi$) dependence of $I_{\pi}$/$I_{\sigma}$ at the $\mathbf{q}=$ \qhalfneg magnetic Bragg peak, for photon energies tuned to the Ni $L_3$ and Dy $M_5$ absorption edges \cite{additional_info}. Incident or scattered photons are shadowed by the sample (prohibiting the experiment) in the angular regions indicated with a grey bar. The data point at $\psi = 90^{\circ}$ was measured in an alternate geometry \cite{Supplementary_Material}.  (c) Temperature dependence of $I_{\pi}$/$I_{\sigma}$ at the Ni $L_3$ edge for $\psi=30^{\circ}$ in magnetic fields of $H = 0$ (solid squares) and $H = 5$ T (open circles) oriented along the [10$\overline{1}$] direction. The integrated magnetic scattering intensity at the Dy $M_5$ edge for $\pi$-polarized light is shown in red along with a fit $\propto 1/T$. The inset shows the Ni $L_3$ ratio $I_{\pi}$/$I_{\sigma}$ at $T=4.4$ K as a function of $H$.   \label{fig:Dy_peak_double_fig_v1}}
\end{figure*}

We now discuss the RXS data which provide insight into the microscopic arrangement of the Dy and Ni magnetic moments in the LaNiO$_3$-DyScO$_3$ superlattices \cite{PhysRevB.73.100409, PhysRevB.77.115138}. Previous Ni $L_3$-edge RXS experiments on a ($2\vert\vert2$) LaNiO$_3$-\dso superlattice grown on DyScO$_3$ revealed a commensurate spiral state with magnetic Bragg reflections at momentum transfer $\mathbf{q}=$ ($\frac{1}{2}$ 0 $\frac{1}{2}$) [$\mathbf{q}=$ ($\frac{1}{4}$ $\frac{1}{4}$ $\frac{1}{4}$) in pseudocubic notation] below $T_{N} \sim 100 $ K  \cite{Frano_SDW_nickelates}. Here we report the observation of a magnetic Bragg reflection at the same momentum transfer $\mathbf{q}$, but for photons tuned to the Dy $M_5$ edge (Fig. \ref{fig:Dy_peak_double_fig_v1}(a)). This reflection is not observed in bulk DyScO$_3$. It appears in the superlattice at low temperatures (Fig. \ref{fig:Dy_peak_double_fig_v1}(c), red diamonds), and exhibits an energy dependence characteristic of resonant scattering from the first order magnetic Bragg reflection of ordered Dy$^{3+}$ moments (inset in Fig. \ref{fig:Dy_peak_double_fig_v1}(a)), with two peaks corresponding approximately to transitions of the type $\Delta M_J = \pm 1$ \cite{PhysRevLett.105.167207, PhysRevB.74.094412}.

The magnetic Bragg reflection is also detectable at the Dy edge for a ($2\vert\vert4$) superlattice (4 u.c. of \dso per bilayer), but no longer for a ($4\vert\vert4$) superlattice. Figure \ref{fig:Dy_peak_double_fig_v1}(a) provides a comparison of rocking curves around momentum transfer \qhalf for the three samples, where the scattered intensity is normalized to the total number of \dso layers. The weaker intensity recorded for the ($2\vert\vert4$) superlattice indicates that the magnetic order is induced only in the interfacial \dso layers. Since superlattices with 4 u.c. of \lno per bilayer do not exhibit antiferromagnetic order in the nickelate layers, the lack of magnetic scattering at the Dy $M_5$ edge for the ($4\vert\vert4$) sample demonstrates that the magnetic order at the Dy sites does not result from confinement of the DyScO$_3$ layers, but is rather induced in proximity to antiferromagnetic LaNiO$_3$ by Ni-Dy exchange interactions across the epitaxial interface.

The direction of the magnetic moments in the superlattice can be inferred from the photon polarization dependence of the RXS cross section. In a magnetic RXS experiment, the scattered intensity is given by
\begin{equation}
I_{\mu \nu} = \vert \sum_{i} e^{\imath\mathbf{q}\cdot\mathbf{r}_i} F_i (E) \; \varepsilon_{\mu}\times\varepsilon^{*}_{\nu}\cdot\mathbf{\hat{m}}_i) \ \vert^2 ,
\label{eqn:magnetic_scattering}
\end{equation}
to first order in the magnetic moment $\mathbf{\hat{m}}_i$ of the ion located at $\mathbf{r}_i$.  $\varepsilon_{\mu(\nu)}$ is the polarization of the incoming (outgoing) light, and $F_i$ is the photon energy ($E$) dependent scattering tensor \cite{Hill_McMorrow_xray_resonant_exchange_scattering}.  By rotating the sample about $\mathbf{q}$, the Bragg condition defined by the phase factor $e^{\imath\mathbf{q}\cdot\mathbf{r}_i}$ in equation \ref{eqn:magnetic_scattering} is preserved, but the scattered intensity is modulated as the projection of $\varepsilon_{\mu}$ onto $\mathbf{\hat{m}}_i$ changes. Rotation about $\mathbf{q}$ is parameterized by the azimuthal angle $\psi$ which is defined independently for each $\mathbf{q}$. For $\mathbf{q}$=\qhalfneg we set $\psi = 0$ when the scattering plane is spanned by \qhalfneg and $(1 \,\overline{1}\,2)$, and the projection of the latter onto $\mathbf{k}_{in}$ is positive. $\psi$ increases for left handed rotation of the crystal about $\mathbf{q}$.

\begin{figure*}[htb]
\includegraphics[width=15.cm]{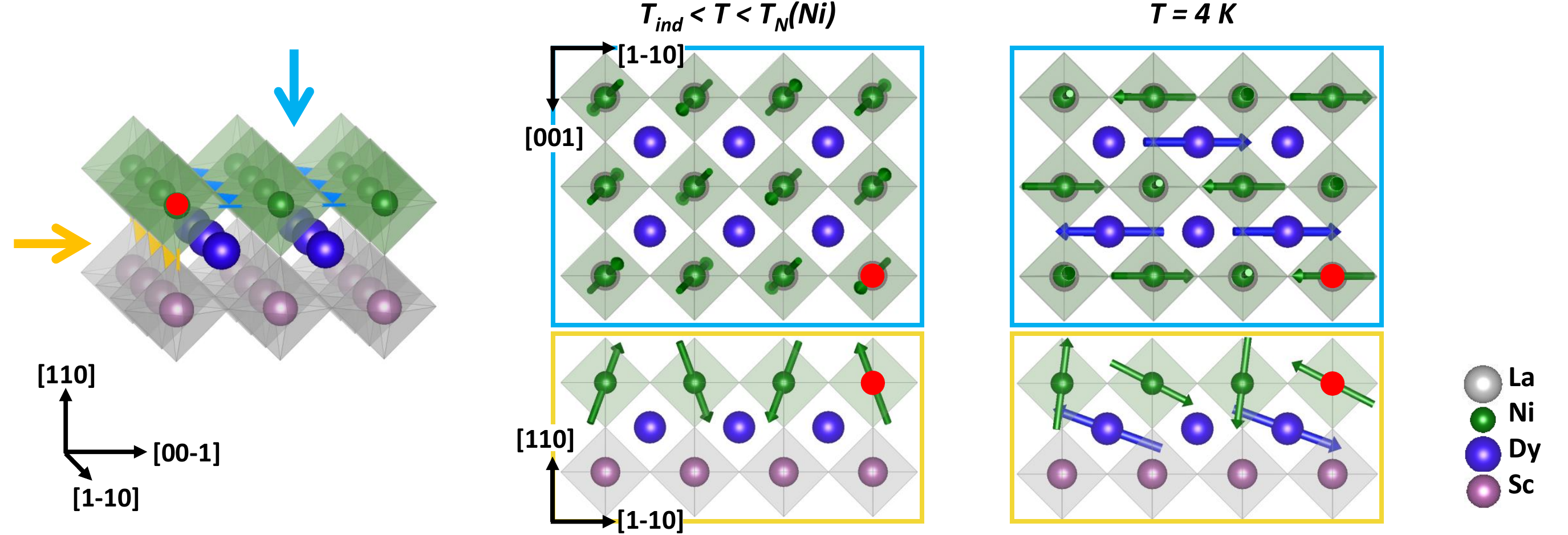}
\caption{A depiction of the magnetic structure at the LaNiO$_3$-DyScO$_3$ interface \cite{second_additional_info} at temperatures above $T_{ind}$ (center) and at $4$ K (right). Since the resonant scattering experiment is not sensitive to a relative phase shift between the Ni and Dy magnetic structures, it is assumed that the Ni-Dy exchange interaction is antiferromagnetic. The lack of an arrow at a Dy site indicates the absence of detected magnetic order. To simplify the sketch, orthorhombic lattice distortions have been omitted, and the lengths of the Dy moments are not to scale. \label{fig:temp_triple_fig}}
\end
{figure*}

We recorded the integrated magnetic intensity (i.e., the area of a rocking curve about the magnetic Bragg peak) at $\mathbf{q}=$\qhalfneg as a function of $\psi$, for both $\pi$ and $\sigma$ polarization of the incoming light (i.e., $\varepsilon_{\mu}$ perpendicular and parallel to the scattering plane, respectively). The polarization of the scattered light is not analyzed, so that $I_{\pi}\equiv I_{\pi\pi}+I_{\pi\sigma}$ and $I_{\sigma}\equiv I_{\sigma\sigma}+I_{\sigma\pi}$. Figure \ref{fig:Dy_peak_double_fig_v1}(b) shows the ratio $I_{\pi}$/$I_{\sigma}$ of integrated intensities measured at the Dy $M_5$ edge together with a model calculation of the $\psi$-dependence of $I_{\pi}$/$I_{\sigma}$ based on a collinear antiferromagnetic model, with the two angles specifying the Dy moment direction as free parameters.
The best agreement was found for a structure in which the Dy moments form an angle of $25^{\circ}\pm4^{\circ}$ with the $Pbnm$ $b$-axis in the $a$-$b$ plane (right panel of Fig. \ref{fig:temp_triple_fig}).

Interestingly, the crystallographic axis along which the ordered dysprosium moments are oriented agrees closely with one of the two Ising axes determined for bulk DyScO$_3$ (Fig. \ref{fig:bulk_DSO_magnetometry}). A magnetic structure that singles out one of the two moment directions of bulk DyScO$_3$ is indeed expected in this situation, because the LaNiO$_3$-DyScO$_3$ interface is oriented such that the two Dy sites become crystallographically inequivalent (Fig. \ref{fig:bulk_DSO_magnetometry}(b)). The RXS experiment cannot distinguish between structures of the form ``up-up-down-down'' and ``up-zero-down-zero'' (or linear combinations thereof). We note, however, that collinear structures with moments on all sites are strongly disfavored by the alternating Ising anisotropy of DyScO$_3$ (Fig. \ref{fig:bulk_DSO_magnetometry}), and that the RXS data rule out any non-collinear structure consistent with the alternating anisotropy. An ``up-zero-down-zero'' sequence of ordered moments, on the other hand, is physically plausible, because bulk DyScO$_3$ is paramagnetic in the temperature range covered by our experiments ($T \geq 4$ K). The antiparallel orientation of next-nearest-neighbor Dy moments in the ``up-zero-down-zero'' structure is well matched to the commensurate spiral state of LaNiO$_3$ (Fig. \ref{fig:temp_triple_fig}). The small or vanishing amplitude of the ordered moment on the other Dy site can then be ascribed to a reduced Ising anisotropy under the altered crystal-field conditions at the interface and/or competition between the Ising anisotropy and the exchange field. Both mechanisms weaken the effective field acting on the Dy moments such that it is insufficient to quench the magnetic fluctuations. Microscopic model calculations are required to quantitatively determine the magnetic Hamiltonian at the interface.

We now turn our attention to the magnetic structure of the Ni sublattice. In previous work on spiral order in tensile-strained PrNiO$_3$ and LaNiO$_3$ films/superlattices, the Ni spins were found to tilt symmetrically  $28^{\circ}$ from the pseudocubic $c$ axis, with in-plane projections lying symmetrically between the pseudocubic $a$- and $b$-axes \cite{Frano_SDW_nickelates}. For $T \geq T_{ind} \sim 18$ K, the Ni $L_3$ edge data on our superlattices agree with this result (red line in Fig. \ref{fig:Dy_peak_double_fig_v1}(b)). Upon cooling below $T_{ind}$, however, we observe pronounced intensity changes of the Ni $L_3$ Bragg reflections, heralding a temperature dependent spin reorientation concomitant with the onset of induced magnetic order in the \dso layers (Fig. \ref{fig:Dy_peak_double_fig_v1}(c)).

Figure \ref{fig:Dy_peak_double_fig_v1}(b) shows that the azimuthal dependence of $I_{\pi}$/$I_{\sigma}$ at the Ni $L_3$ edge for $T < T_{ind}$ differs from the one for $T \geq T_{ind}$, confirming a reorientation of the Ni spins. The blue line in Fig. \ref{fig:Dy_peak_double_fig_v1}(b) represents the results of a model calculation for the magnetic structure depicted in Fig. \ref{fig:temp_triple_fig}, which yields the best agreement with the experimental data collected at 4 K \cite{Supplementary_Material}. The interaction across the interface generates two inequivalent Ni sites, depending on whether the two adjacent ordered Dy moments are (1) parallel or (2) antiparallel to one another (see Fig. \ref{fig:temp_triple_fig}, right panel). Ni spins which see two parallel Dy moments reorient almost exactly antiparallel to these moments (in the $a$-$b$ plane, 18$^{\circ} \pm 4^{\circ}$ from [010]), thereby gaining Ni-Dy exchange energy at the expense of the magneto-crystalline anisotropy energy resulting from the weak intra-atomic spin-orbit interaction of Ni. For Ni spins which see two antiparallel Dy moments, the Ni-Dy coupling is frustrated, and their orientation is determined exclusively by Ni-Ni exchange interactions in the spiral state (almost directly perpendicular to the surface at 8$^{\circ} \pm 5^{\circ}$ from [110] and rotated in the sample plane 28$^{\circ} \pm 20^{\circ}$ from [$\overline{1}$10] towards [001]).

Finally we address the response of the system to an external magnetic field $H$. In x-ray magnetic circular dichroism (XMCD) measurements at the Dy $M_5$ edge for $H$ along the [1 0 $\overline{1}$] direction at $T = 4.4$ K, we observed a magnetic field-induced net Dy moment with 88$\%$ saturation at  $H =3$ T (data not shown). It is expected that such a field induced breaking of the \qhalf Dy magnetic structure renders the Ni-Dy exchange interaction ineffective, and that the temperature-induced reorientation of the spin spiral is thus reversed for sufficiently high $H$. RXS experiments at the Ni $L_3$ edge indeed reveal the expected $H$-induced spin-flop transition. The inset of Fig. \ref{fig:Dy_peak_double_fig_v1}(c) shows $I_{\pi}$/$I_{\sigma}$ at the Ni $L_3$ edge for $\psi=30^{\circ}$ as a function of applied field. The reorientation transition manifests itself as an upturn in $I_{\pi}$/$I_{\sigma}$ that saturates around 3 T, consistent with the Dy XMCD data. The $H$-induced spin reorientation is fully reversed upon removal of the field.

In summary, our comprehensive Dy $M_5$ and Ni $L_3$ edge RXS experiments have revealed an unusual magnetic ground state generated by $3d-4f$ exchange interactions across LaNiO$_3$-DyScO$_3$ interfaces. Whereas the Ni spin system imposes the periodicity of its magnetic order on the interfacial Dy moments, these moments in turn, rigid due to their large magneto-crystalline anisotropy, pin the Ni spins and induce a reorientation transition in the two-unit-cell thick LaNiO$_3$ layer. With their 10 $\mu_B$ amplitude, the Dy moments provide ``anchoring points'' for external magnetic fields, which enable field-induced transitions for moderate field strengths. In contrast, applying magnetic fields of up to 5 T to LaNiO$_3$-LaAlO$_3$ superlattices, where no rare earth magnetic moment is present, induces only a slight canting of the Ni spin spiral \cite{Wu_private_communication}. The insertion of layers containing ``anchor spins'' could enable the control of the relative orientation of antiferromagnetic structures in spintronic devices and thus facilitate previously infeasible spin transport experiments.

Related effects have been investigated in the bulk multiferroic manganites TbMnO$_3$ and DyMnO$_3$, where magnetic-field-induced spin reorientations induce an electric polarization flop \cite{kimura_TbMnO3_nature, Kimura_phase_diagrams, Mochizuki_2009_PRB, Cuartero_2015_PRB, Prokhnenko_2007_PRL_TbMnO3, Prokhnenko_DyMnO3, Kenzelmann_TbMnO3_PRL_2005}. We emphasize, however, that the superlattice approach we have introduced here is quite distinct from the effect of chemical substitution in perovskites with electronically active transition metals. For instance, replacing La in LaNiO$_3$ with the smaller Dy ion reduces the Ni-O-Ni bond angle and the electronic band width, and thus drives a phase transition into a charge-ordered insulating state. In contrast, the lattice structure and chemical bonding pattern in our superlattices are only minimally disturbed by the adjacent DyScO$_3$ layers, and LaNiO$_3$ can maintain its high electrical conductivity. Tailored $3d-4f$ exchange interactions in metal-oxide heterostructures and devices thus open up new perspectives for the manipulation of spin structures by external magnetic fields, complementary to the manipulation of the carrier density by gate electric fields.

We thank the German Science Foundation (DFG) for financial support under grant No. SFB/TRR 80.


\begin{thebibliography}{33}
\bibitem [{add()}]{additional_info}%
  \BibitemOpen
  \bibinfo {note} {In order to access the momentum transfer ($\frac{1}{2}$ 0
  $\overline{\frac{1}{2}}$) in the horizontal scattering geometry, the samples
  were mounted on a wedge-shaped holder with an angle of 35$^{\circ}$ with
  respect to the scattering plane.}\BibitemShut {Stop}%
\bibitem [{sec()}]{second_additional_info}%
  \BibitemOpen
  \bibinfo {note} {The Dy-NiO$_2$ interface termination implied in the drawing
  is not necessarily present throughout the sample, but other terminations are
  magnetically inactive and hence not relevant for our study.}\BibitemShut
  {Stop}%
\end{thebibliography}

%

\end{document}